\documentclass[a4paper,12pt,reqno,superscriptaddress,nofootinbib]{revtex4}
\usepackage[centertags]{amsmath}
\usepackage{amsfonts}
\usepackage{amssymb}
\usepackage{amsthm}
\usepackage{newlfont}
\usepackage{stmaryrd}
\usepackage{mathrsfs}
\usepackage{euscript}
\usepackage{siunitx}
\usepackage{physics}
\usepackage{algorithm2e}
\usepackage{graphicx,subcaption}
\usepackage{enumitem}
\usepackage{natbib}
\usepackage{graphicx}
\usepackage{color}
\usepackage{floatrow}
\usepackage{caption}
\usepackage{hyperref}
\usepackage{hhline}
\usepackage{hyperref}
\usepackage{cleveref}
\usepackage{import}

\theoremstyle{plain}

\theoremstyle{definition}

\theoremstyle{remark}

\numberwithin{equation}{section}

\let\ve=\varepsilon

\newcommand{\opunit}{\text{1}\kern-0.22em\text{l}}

\DeclareMathAlphabet{\mathpzc}{OT1}{pzc}{m}{it}

\newcommand{\id}{\textrm{d}}

\usepackage{floatrow}
\usepackage{caption}
\usepackage{color,soul}

\DeclareCaptionJustification{justified}{\leftskip=0pt \rightskip=0pt \parfillskip=0pt plus 1fil}
\begin{document}

\title{Specific heat of the driven Curie--Weiss model}

\author{Elena Rufeil Fiori$^{1,2}$, Christian Maes$^1$ and Robbe Vidts$^1$\\ {\it$^1$Department of Physics and Astronomy, KU Leuven, Belgium.}\\{\it$^2$Facultad de Matemática, Astronomía, Física y Computación, CONICET--Universidad Nacional de C\'ordoba, Argentina.}}


\begin{abstract}
Applying a time--periodic magnetic field to the standard ferromagnetic Curie--Weiss model brings the spin system in a steady out--of--equilibrium condition. We recall how the hysteresis gets influenced by the amplitude and the frequency of that field, and how an amplitude- and frequency- dependent (dynamical) critical temperature can be discerned. The dissipated power measures the area of the hysteresis loop and changes with temperature. The excess heat determines a nonequilibrium specific heat giving the quasistatic response. We compute that specific heat, which appears to diverge at the critical temperature, quite different from the equilibrium case.  
\end{abstract}

\maketitle


\section{Introduction}

Nonequilibrium calorimetry started with the macabre experiments of Lavoisier--Laplace (1782).  It has fueled the field of bioenergetics which turned to its molecular basis, to remain very much alive for understanding cellular powerhouses.  In hard condensed matter, the measurement of nonequilibrium heat capacities was pioneered in a series of papers by del Cerro {\it et al.}, \cite{cerro89, cera, cerro96}.
These experiments measured heat fluxes on LATGS single crystals (triglyine sulfate grown with small amounts of L--alanine), a ferroelectric material that shows electrocaloric effects. Dissipated power is created there by a time--periodic electric field.
The present paper gives a theoretical analysis for the magnetic counterpart.\\
Driven mean--field systems make good ground for starting a study of spatially extended nonequilibria, and to reveal how their thermal properties and criticality differ from their equilibrium versions.
The Curie--Weiss model is a mean--field version of the Ising model where thermodynamic properties can be calculated explicitly when the number of spins tends to infinity. For nonequilibrium version, we apply a magnetic field that varies periodically with a given frequency $\omega_0$. The magnetization reaches a steady nonequilibrium condition, and depending on temperature and $\omega_0$, a hysteresis curve is drawn in the magnetization {\it versus} magnetic field plane. The area is a measure of the dissipated heat, which changes under small temperature variations giving rise to the notion of excess heat. The corresponding (nonequilibrium) heat capacity summarizes the quasistatic response of that excess heat to the temperature variations, \cite{epl}.\\
The computational method essentially follows the experimental procedure for measuring such a heat capacity. We refer to \cite{cera} for the experimental example, and to \cite{TLSpre2024} for an elementary exposition and (theoretical) example of the essential steps. There and in contrast to previous work in e.g. \cite{epl, cal, simon, activePritha, jchemphys, prithaarray} the (unperturbed) system under consideration is periodic in time and the source of nonequilibrium is the time--dependence of the Hamiltonian.\\

We see such studies as in the present paper primarily as the beginning of thermal physics for spatially extended nonequilibrium systems. In particular, the heat capacity (or, specific heat when considered per spin) reveals information about the phase diagram of the periodically--driven Curie--Weiss model, as it did for a two--temperature version of the same model, \cite{aaron2024}. The critical temperature decreases and the specific heat is divergent there in contrast with the mean--field case in equilibrium.\\

In the next section, we recall the dynamics of the time--dependent Curie--Weiss model and how to obtain the macroscopic evolution equation for the magnetization. In particular, it reproduces the hysteresis loop, along which the system steadily dissipates heat. In Section \ref{power}, we define and compute the ingredients such as the Joule heating and the dissipated power that will lead to the heat capacity. We cannot compute it exactly but there remain two ways to proceed theoretically, via numerical methods (Section \ref{num}), or via simulation (Section \ref{fss}) of the spin--flip dynamics. We do both and obtain the same heat capacity curves as a function of temperature and driving amplitude. 

\section{Driven Curie--Weiss model}\label{tdh}

For general references to the theoretical study of the equilibrium Curie--Weiss model and for its description of the behavior of magnetic materials, see e.g. \cite{bookcw, revcw}.\\

Consider a numbered collection of spins $\sigma = (\sigma_1, \sigma_2, \dots, \sigma_N)$ where $\sigma_i = \pm1$ and $N$ is large.
The magnetization as a function of the spin configuration $\sigma$ is 
\begin{equation}\label{mag}
    m^N(\sigma) = \frac{1}{N}\sum_{i=1}^{N} \sigma_i
\end{equation}
and the energy only depends on that magnetization, as
\begin{equation}
    E^N(\sigma) = -{\cal E} N\psi(m^N(\sigma)) \qquad \text{with} \qquad \psi(m) = \frac{m^2}{2} + h\,m
\label{eq_energy}
\end{equation}
The ${\cal E}$ gives the energy units for $E^N$, which we will ignore and take equal to one.
The parameter $h = h_t$ corresponds to the magnetic field and is time--dependent, as will be outlined below and which asks for a dynamics.\\

We imagine that the spins are in contact with a thermal bath at inverse temperature $\beta$ ($k_B=1$). 
The dynamics follows the transitions of a kinetic Ising model, where the only allowed transitions are between $\sigma = (\sigma_1,\dots, \sigma_k,\dots, \sigma_N)$ and $\sigma^k = (\sigma_1,\dots, -\sigma_k,\dots, \sigma_N)$, for some $k \in \{1,2,...,N\}$. That is to say, only one spin flip at a time is allowed.
Then, the (driven or not) Curie--Weiss dynamics is a Markov process $\sigma_t$ on $K_N=\{-1,+1\}^N$ for which we specify the transition rate for the spin flip $\sigma \rightarrow \sigma^k$ to be
\begin{equation}\label{tr}
    c^N(\sigma,k) = \nu(k,\sigma)\, \frac 1{1 + e^{\beta\left(E^N(\sigma^k)-E^N(\sigma)\right)}}
\end{equation}
We have chosen to highlight a form for rates that remain bounded at vanishing temperatures (as is physically most reasonable). The symmetric prefactor $\nu(k,\sigma) = \nu(k,\sigma^k)>0$ may however also be chosen to depend on temperature and energy, and allows other transition rates that are all compatible with instantaneous detailed balance: 
\begin{equation}
\frac{c^N(\sigma,k)}{c^N(\sigma^k,k) } = e^{\beta\left(E^N(\sigma)-E^N(\sigma^k)\right)}
\end{equation}
At any rate, in the spirit of mean--field, it is natural to take $\nu(k,\sigma) = \nu(k,m^N(\sigma))$  a function of the magnetization.\\
When the magnetic field is constant in time, there is a unique stationary distribution for each $N$; it is the equilibrium distribution $\sim \exp[-\beta E^N(\sigma)]$ for energy \ref{eq_energy} with inverse temperature $\beta$.  The (equilibrium) heat capacity is the derivative with respect to the temperature, $\beta^{-1}$, of the mean energy in that ensemble. In the thermodynamic limit $N\uparrow \infty$ there is a phase transition at $\beta=1$, where the specific heat shows a discontinuity; see \cite{bookcw, revcw}.\\ 

In the present paper, we  suppose  work is being done on the magnetic system by periodic variation of the magnetic field, applying
\begin{equation}\label{magnf}
    h_t = h_0\, \sin \omega_0 t
\end{equation}
with amplitude $h_0$ and frequency $\omega_0$. That time--dependence makes $\nu$ and $E^N$ in \eqref{tr} time--dependent to create a time--periodic kinetic Curie--Weiss model. In doing so, we create a steady nonequilibrium spin system, characterized by the inverse temperature $\beta$ in the environment and the driving parameters $(h_0,\omega_0)$.\\
We are interested in its thermal response. As a stimulus, we consider a small variation of the temperature: we slowly vary the bath temperature $\beta^{-1}$, with
\begin{equation}
\label{timebeta}
\beta_t = \beta \,(1+ \varepsilon \sin(\omega_\text{B} t))
\end{equation}
using a small frequency $\omega_\text{B} \ll \omega_0$ and small amplitude $\ve$, allowing the quasistatic regime of thermal response.\\
As a consequence, there arises a dissipated power ${\cal P}(t)$ in which the time--dependence \eqref{timebeta} of the temperature enters perturbatively,
\begin{equation}
{\cal P}(t) = {\cal P}^0(t) + \ve \,{\cal P}^1(t) + O(\ve^2) \label{disp}
\end{equation}
The driven spin system at fixed $\beta$, {\it i.e.}, for $\ve=0$ but keeping \eqref{magnf}, is dissipating heat with a power ${\cal P}^0(t)$ to the thermal bath. After some time, ${\cal P}^0(t)$ is periodic with the same frequency $\omega_0$ of the magnetic field. However, ${\cal P}^1(t)$ in \eqref{disp} depends on $\omega_\text{B}$, and the total dissipated power ${\cal P}(t)$ need not be periodic, except when $\omega_0$ and $\omega_\text{B}$ are commensurable.\\
These fluxes can be measured, {\it e.g.}, via a thermopile that converts heat into electric work \cite{cerro89, cera, cerro96}, but here we want to compute them for the above Curie--Weiss setup.

\section{Stochastic and macroscopic evolution}\label{power}

The simplification of the mean--field modeling is that both the statics and the dynamics get closed descriptions in terms of the magnetization alone. For every $N$, the dynamics \eqref{tr} on the spins induces a Markov jump process for the magnetization variable $m \in \{-1,-1+ 2/N,\ldots, 1-2/N,1\}$, with (time--dependent) transition rates 
\begin{equation}
\begin{aligned}
   W_t(m\rightarrow m+\frac{2}{N}) & = N(1-m)\,\nu_t(m) \,\frac{1}{1+e^{\beta_t\,N\left[\psi_t(m) - \psi_t\left(m+\frac{2}{N}\right)\right]}} \label{eq:fin_trans_rate} \\ 
  W_t(m+ \frac{2}{N}\rightarrow m) & = N(1+m + \frac{2}{N}) \,\nu_t(m)\,\frac{1}{1+e^{\beta_t\,N\left[\psi_t(m+\frac{2}{N}) - \psi_t\left(m\right)\right]}}  
\end{aligned}  
\end{equation}
We have made explicit the time--dependence of temperature (like in \eqref{timebeta}) and magnetic field (like in \eqref{magnf}) in $\psi_t(m) = m^2/2 + h_t\,m$. The prefactor $N$ takes into account that changes of the magnetization are of the order $1/N$, and $\nu_t(m)> 0$ parametrizes the rate with a magnetization--dependent frequency while keeping instantaneous detailed balance with respect to the Landau functional
\begin{equation}
\cal L_t(m) = \log \frac{(\frac{1+m}{2}N)!(\frac{1-m}{2}N)!}{N!} - \beta_t N\,\psi_t(m),\quad\,  \frac{ W_t(m\rightarrow m+\frac{2}{N})}{ W_t(m+ \frac{2}{N}\rightarrow m)} = e^{\cal L_t(m)- \cal L_t(m+\frac{2}{N})}
\end{equation}

\begin{figure}[hbt]
\centering
    \begin{subfigure}[hbt]{0.48\textwidth}
        \centering
        \includegraphics[width=\textwidth]{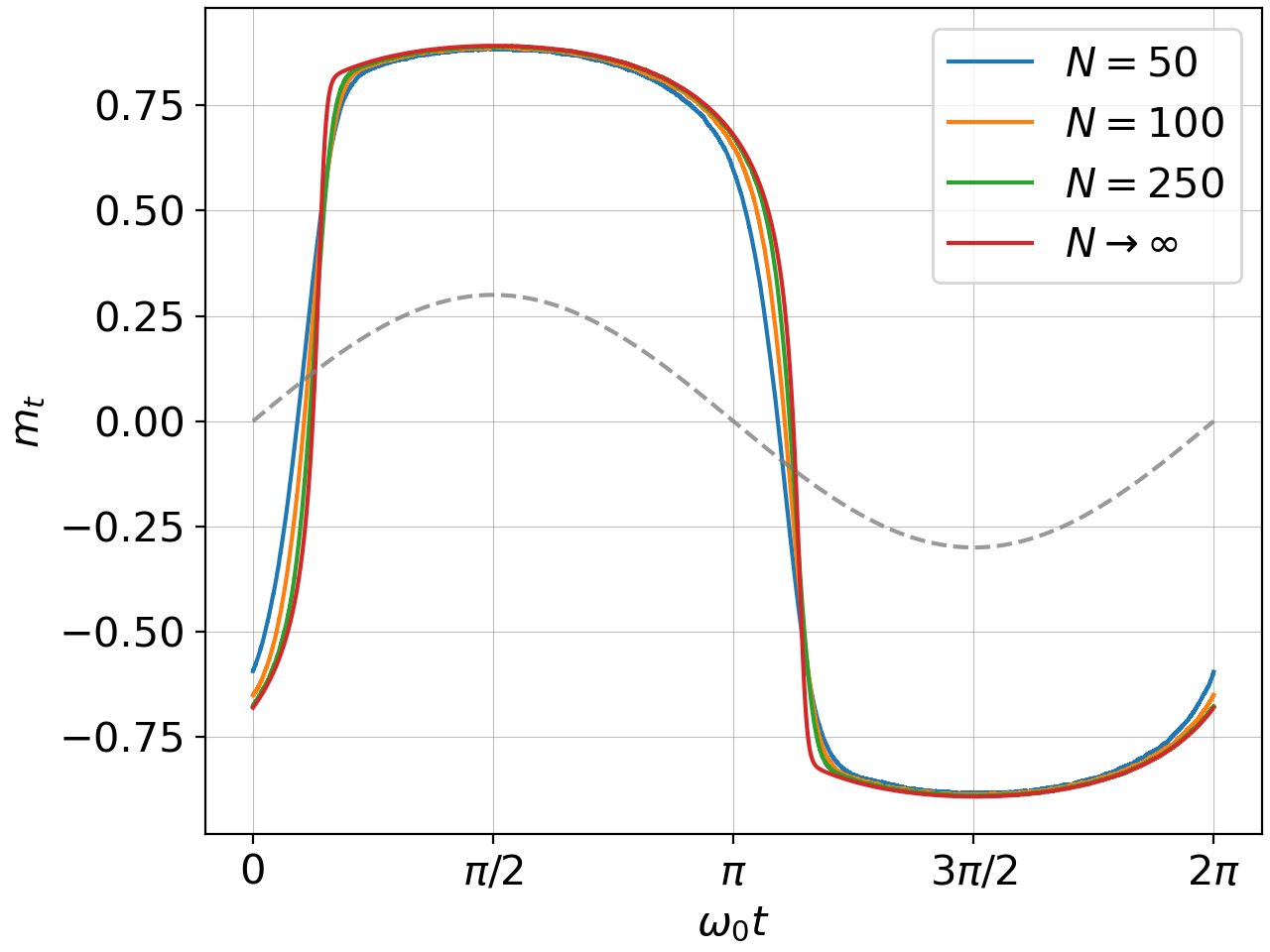}
        \caption{}
    \end{subfigure}
    \hfill
    \begin{subfigure}[hbt]{0.48\textwidth}
        \centering
        \includegraphics[width=\textwidth]{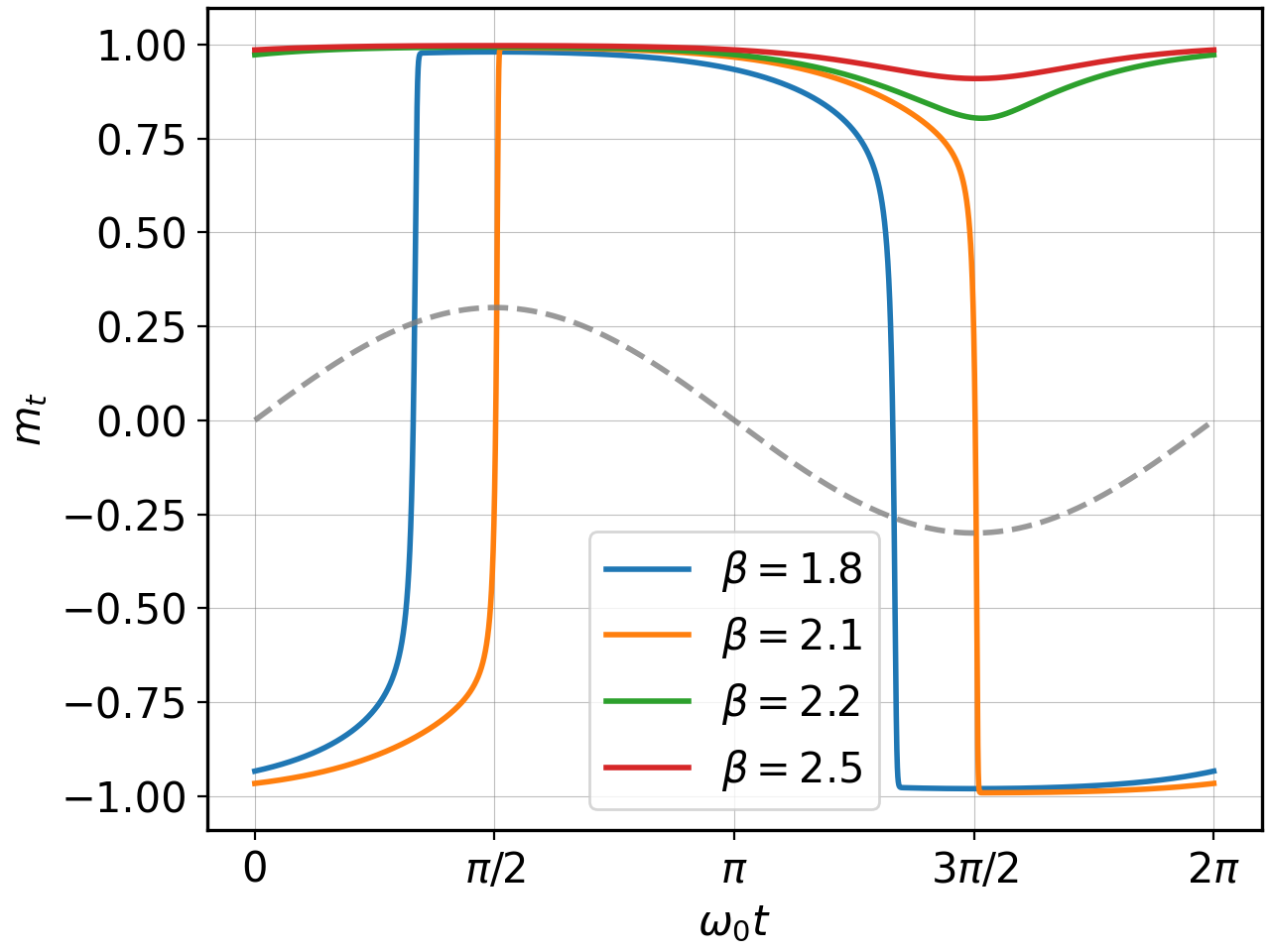}
        \caption{}
    \end{subfigure}
    \caption{Average magnetization as a function of time, with unbounded rates \eqref{bound}, (a) for a system with $50, 100$ and $250$ spins (simulations) and in the $N\rightarrow \infty$ limit (macroscopic evolution). External parameters are: $h_0=0.3$, $\beta = 1.2$, $\omega_0 = 0.02$. The magnetization for the finite spin system was averaged over 5000 runs. As a reference, the magnetic field is plotted in dashed lines. (b) Magnetization for the macroscopic evolution with $h_0=0.3$ and $\omega_0 = 0.02$ for different inverse temperatures $\beta$.  Note how  the system cannot explore both solutions (green and red lines in (b)) for $\beta>\beta_c(h_0=0.3,\omega_0=0.02) \simeq 2.15$.}
    \label{fig_fin_mag}
\end{figure}

In the limit $N\uparrow \infty$, there appears a (deterministic) equation for the (macroscopic) magnetization $m_t$.
We take the large $N-$limit 
in the kinetic equation
\begin{equation}
\frac{\id m_t}{\id t} =  \frac{2}{N}\;\left[W_t(m\rightarrow m+\frac{2}{N}) -  W_t(m\rightarrow m-\frac{2}{N})\right]
\end{equation}
to get the macroscopic evolution equation
\begin{equation}\label{neqCW}
    \frac{\id m_t}{\id t} = 2\nu_t(m_t)\,\left[\tanh \beta_t\psi_t'(m_t) - m_t\right]
\end{equation}
where $\psi_t'(m) = m + h_t$, and $\nu_t(m)>0$ can be chosen freely.\\
The kinetic prefactor $\nu_t$ changes the dynamics. In what follows we refer to bounded {\it versus} unbounded rates, respectively by choosing
\begin{equation}\label{bound}
\nu_t(m) = \frac{1}{2} \quad \text{bounded},\qquad\qquad \nu_t =
\frac{1}{2} \cosh \beta_t\psi'_t(m)\quad \text{unbounded}
\end{equation}

The above leaves us two computational approaches. We either make the stochastic updating following \eqref{eq:fin_trans_rate}, which requires a simulation, or we proceed with the numerical study of the macroscopic evolution \eqref{neqCW}. As a first illustration, we refer to Figs.~\ref{fig_fin_mag}. We see and compare the magnetization from the simulations and from the macroscopic evolution in Fig.~\ref{fig_fin_mag}(a).  Fig.~\ref{fig_fin_mag}(b) gives the magnetization from the macroscopic eveolution for different temperatures.\\
That time-dependent magnetization is used to compute the dissipated power ${\cal P}(t)$ in \eqref{disp}. The latter is obtained from the energy \ref{eq_energy}, by identifying the heat and work contribution in its time--derivative, obtaining
\begin{equation}
\label{eq_P}
    {\cal P}(t) = (m_t+h_t) \frac{\id m_t}{\id t}
\end{equation}
determined by the (variation in the) magnetization $m_t$. As we are interested in linear thermal response, using the temperature variation \eqref{timebeta}, we make an expansion in $\ve$ of the magnetization, $m_t =m_t^0 + \varepsilon \, m_t^1 + O(\ve^2)$, to work with
\begin{equation}
\begin{aligned}
\label{eq_P1}
   {\cal P}^0(t) & = (m_t^0 + h_t)\frac{\id m_t^0}{\id t} \\
   {\cal P}^1(t) & = m_t^1 \frac{\id m_t^0}{\id t} + (m_t^0+h_t) \frac{\id m_t^1}{\id t} 
\end{aligned}
\end{equation}
All of that can be numerically evaluated in Section \ref{num} starting from \eqref{neqCW}, or we can use simulations for a large finite system using \eqref{eq:fin_trans_rate}, which we consider in Section \ref{fss}.

\section{Dynamical hysteresis and criticality}\label{hyst}

In a magnetic spin system, the ramping of the magnetic field triggers spin flips which can lead to avalanches of spin domains changing sign. That can be studied in detail for the Curie--Weiss model, \cite{chat,chak,lak}. 
By eliminating time from the solution $m_t$ of \eqref{neqCW}, we obtain a parametric plot $(m,h)$, and we may see the occurrence of a hysteresis loop depending on $\beta, h_0$ and $\omega_0$.  

\begin{figure}[hbt]
    \centering
    \begin{subfigure}[hbt]{0.48\textwidth}
        \centering
        \includegraphics[width=\textwidth]{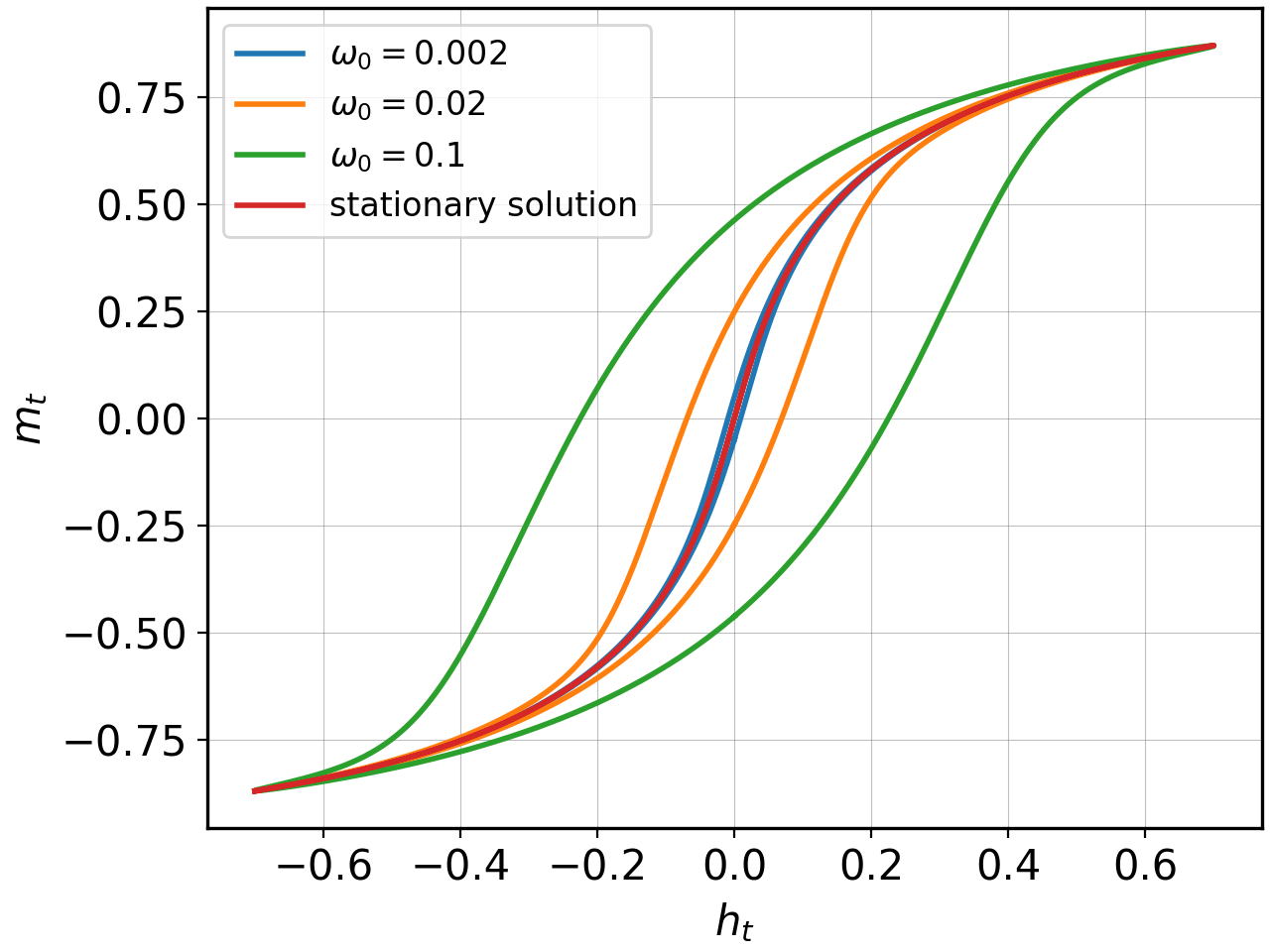}
        \caption{$\beta=0.85$}
    \end{subfigure}
    \hfill
    \begin{subfigure}[hbt]{0.48\textwidth}
        \centering
        \includegraphics[width=\textwidth]{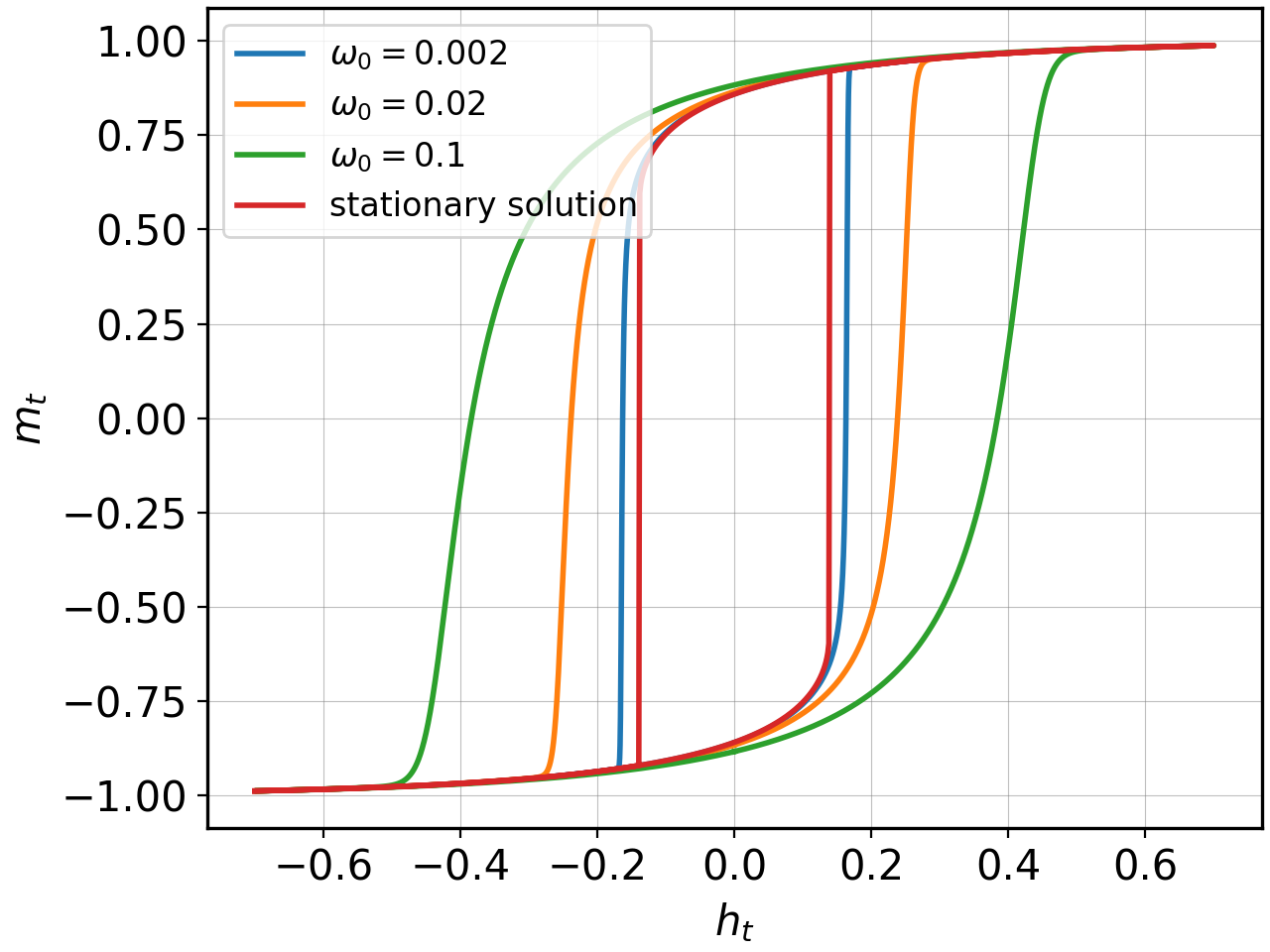}
        \caption{$\beta=1.5$}
    \end{subfigure}
    \caption{Hysteresis loops for two different inverse temperatures $\beta$, obtained from the macroscopic evolution with $h_0=0.7$, with unbounded rates \eqref{bound}}
\label{fig_loops}
\end{figure}

\begin{figure}[hbt]
    \begin{subfigure}[hbt]{0.48\textwidth}
        \centering
        \includegraphics[width=\textwidth]{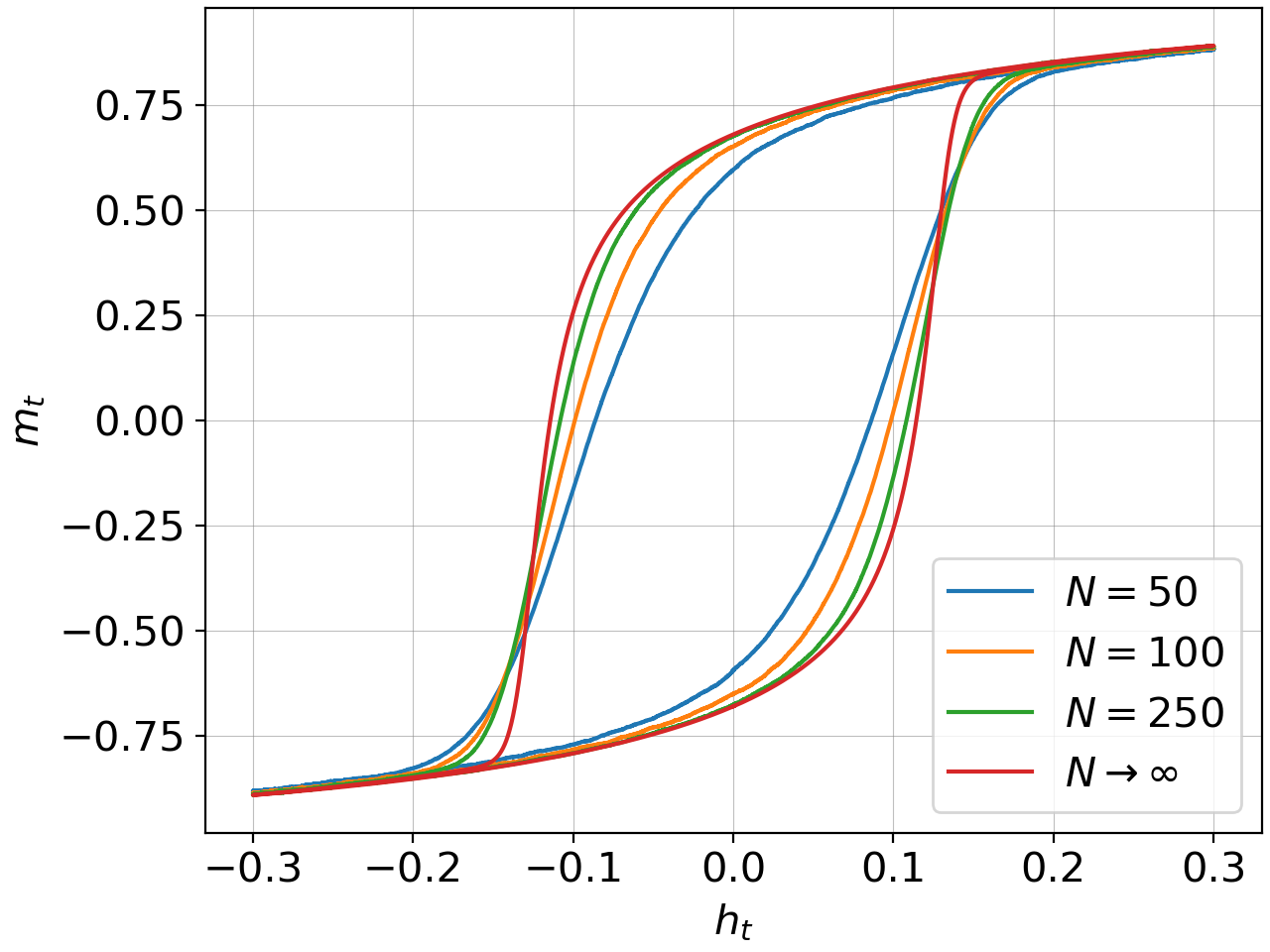}
        \caption{}
    \end{subfigure}
    \hfill
    \begin{subfigure}[hbt]{0.48\textwidth}
        \centering
        \includegraphics[width=\textwidth]{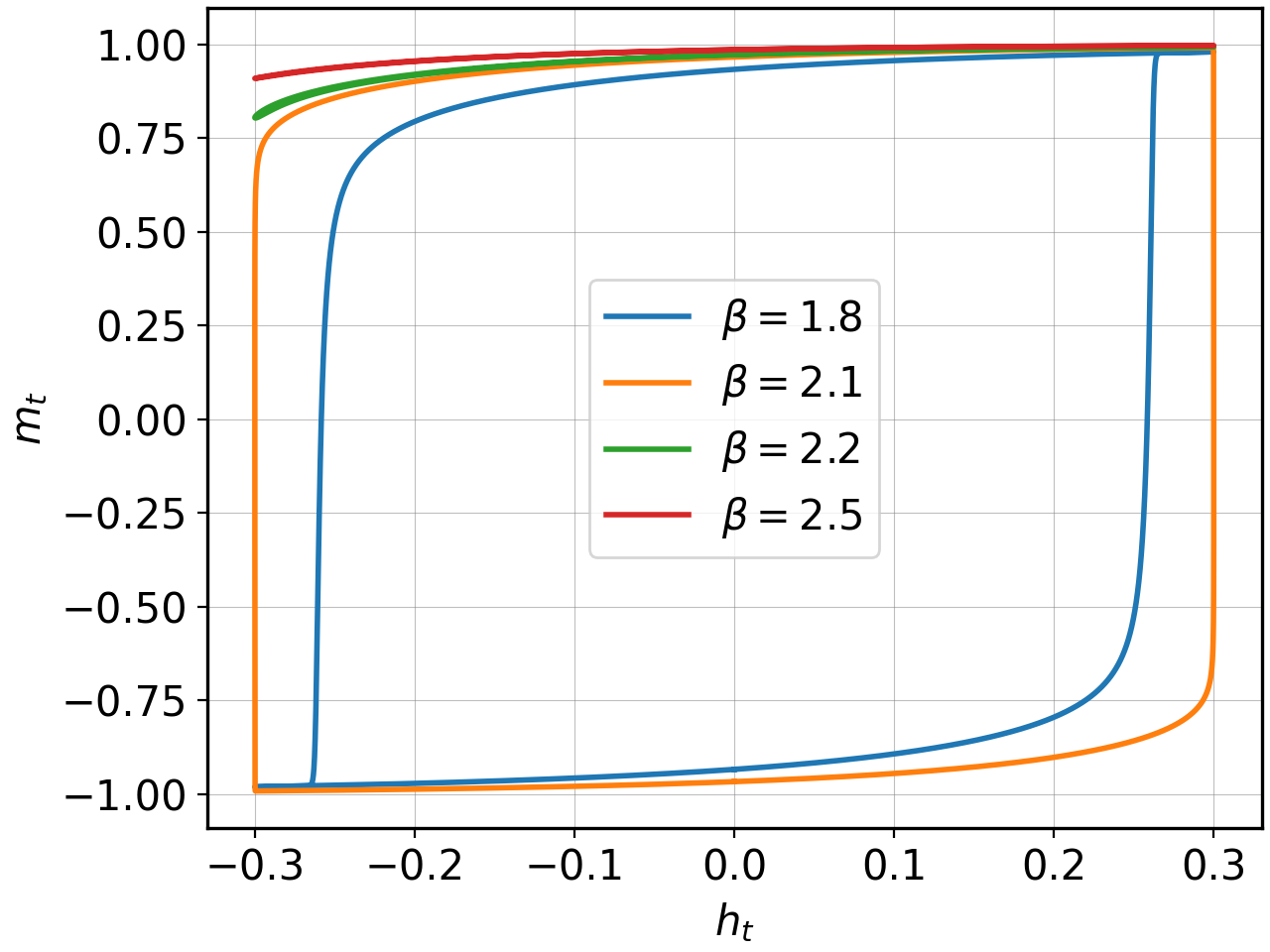}
        \caption{}
    \end{subfigure}
    \caption{Hysteresis loops, corresponding respectively to the situations in Figs.\ref{fig_fin_mag}, (a) and (b), with unbounded rates \eqref{bound}. }
\label{fig_loop_temp}
\end{figure}

\begin{figure}[hbt]
    \begin{subfigure}[hbt]{0.48\textwidth}
        \centering
        \includegraphics[width=\textwidth]{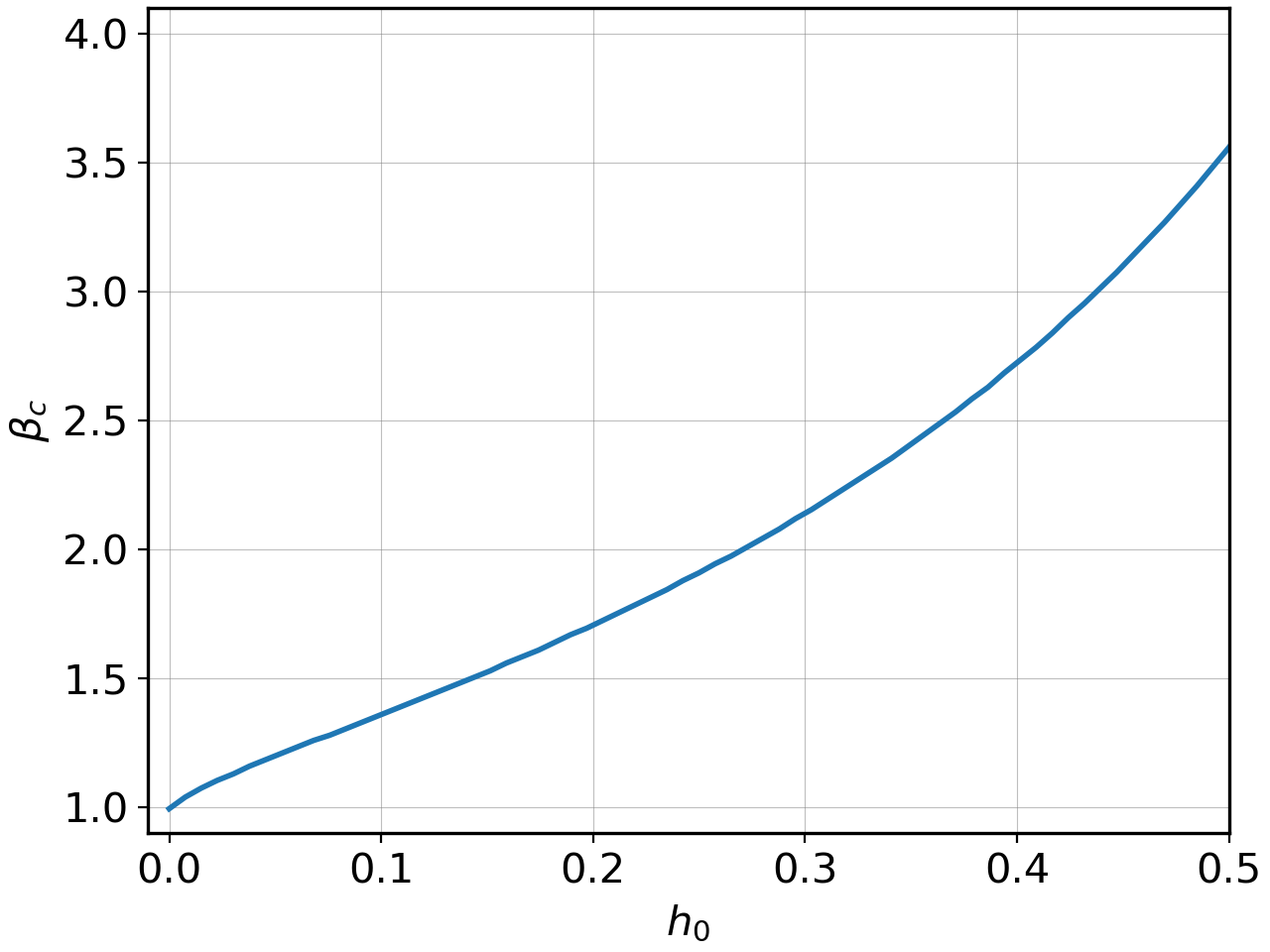}
        \caption{}
    \end{subfigure}
    \hfill
    \begin{subfigure}[hbt]{0.48\textwidth}
        \centering
        \includegraphics[width=\textwidth]{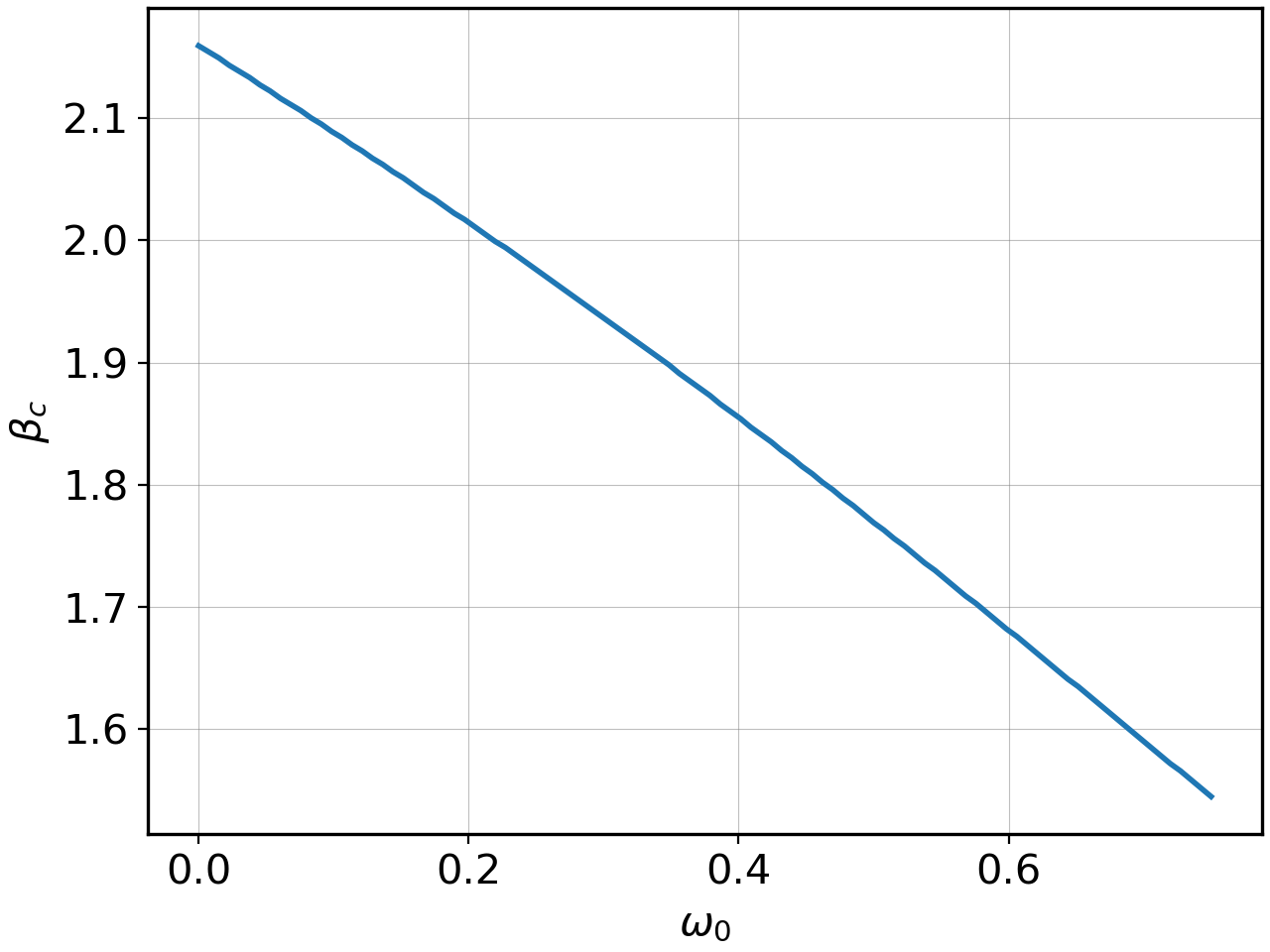}
        \caption{}
    \end{subfigure}
    \caption{The critical temperature $\beta_{c}$ (a) at constant $\omega_0=0.02$, (b) at constant $h_0=0.3$. In both, there is a hysteresis loop (and a change of sign of magnetization) below the line.
    }\label{fig_beta_star}
\end{figure}

The details obviously depend on the values of the amplitude $h_0$ and the frequency $\omega_0$.
The dynamical analysis is a standard topic in time--dependent Landau theory and we restrict ourselves here to verifying the expected behavior. \cite{gal,rieg}.\\
We start with Figs.~\ref{fig_loops}, showing how $\omega_0$ affects the dynamics for two different inverse temperatures.  For $\beta=0.85$ (paramagnetic regime at $h=0$) Fig.~\ref{fig_loops}(a)  shows hysteresis with a narrower loop compared to $\beta=1.5$ in  Fig.~\ref{fig_loops}(b). The hysteresis loop becomes wider for bigger $\omega_0$.  \\ 
When the temperature gets too low (depending on $(h_0,\omega_0)$) the system transitions from exhibiting a hysteresis loop to being trapped in one of the two possible nonzero solutions.  That is illustrated in Fig.~\ref{fig_loop_temp}. 
 In Fig.~\ref{fig_loop_temp}(b) we see the breakdown of hysteresis loop for sufficiently low temperatures.\\
There is indeed a (dynamical) critical temperature $\beta_{c}>1$, where the magnetic field is not strong enough for the magnetization to go from one solution to the other, and it gets trapped in one of them. That critical  temperature depends strongly on $h_0$ and is slowly varying with $\omega_0$ at least when $\omega_0$ is small enough; see \cite{lak}. Fig.~\ref{fig_beta_star} shows the dependence of the critical inverse temperature on the parameters. 

\begin{figure}[hbt]
    \begin{subfigure}[hbt]{0.48\textwidth}
        \centering
        \includegraphics[width=\textwidth]{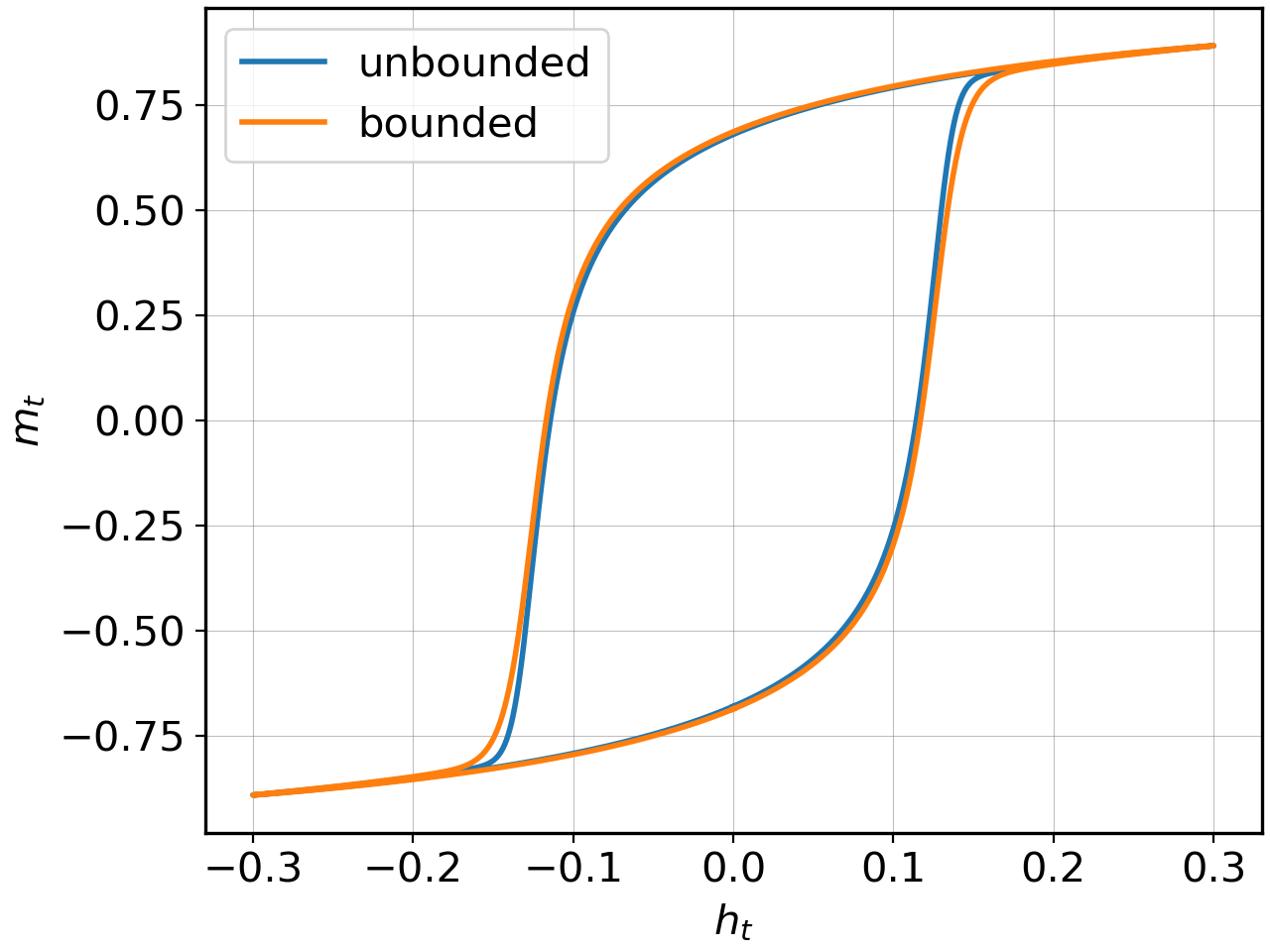}
        \caption{$\beta=1.2$}
    \end{subfigure}
    \hfill
    \begin{subfigure}[hbt]{0.48\textwidth}
        \centering
        \includegraphics[width=\textwidth]{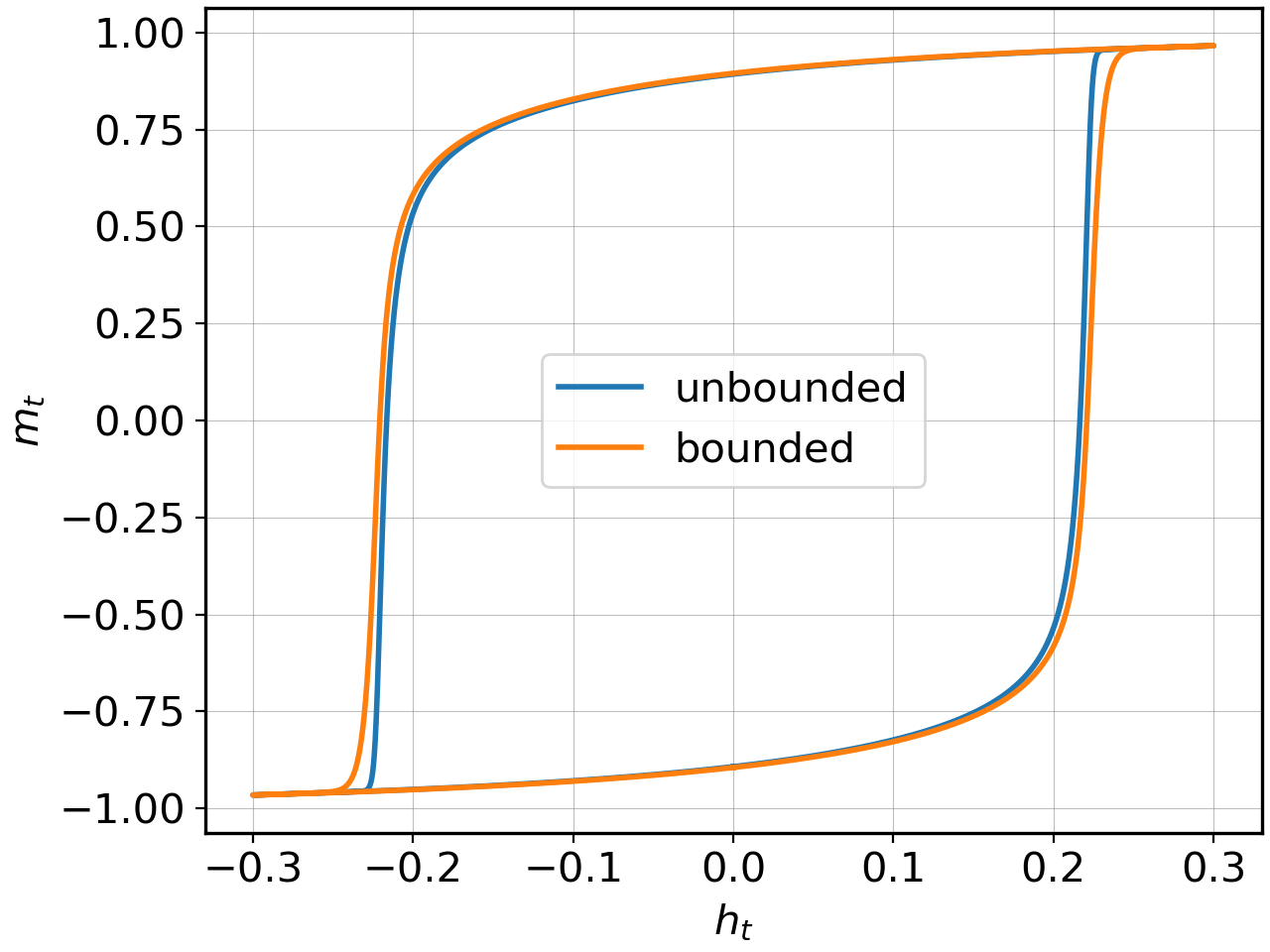}
        \caption{$\beta = 1.6$}
    \end{subfigure}
    \caption{Hysteresis loops for $h_0=0.3$ and $\omega_0 = 0.02$ at two different temperatures, corresponding to bounded and unbounded rates \eqref{bound} for the macroscopic evolution. At higher temperatures the difference between bounded and unbounded rates rapidly becomes invisible.}  
    \label{bofi}
\end{figure}

One must realize that the discussed phenomenon depends on kinetic details, and is therefore called dynamical. As an illustration, we see in Fig.\ref{bofi} how the loop (slightly depends) on the choice of bounded {\it versus} unbounded rates \eqref{bound}.

Possible experimental evidence for the above dynamical phase transition is discussed in \cite{jia,lak}.
What we find in the present paper is that this (dynamical) critical temperature is detected by a divergence of the heat capacity, as we show in the next section.

\section{Heat Capacity}

We continue with the thermal response as introduced by the equations \eqref{eq_P}--\eqref{eq_P1}.
Following the steps of AC--calorimetry \cite{cal} for stationary nonequilibrium systems and in Ref.\cite{dio, TLSpre2024} for steady nonequilibrium systems, the heat capacity $C=C(\beta, \omega_0,\ldots)$ is obtained from the out--of--phase component of minus the excess heat current ${\cal P}^1$: 
\begin{equation}
\label{eq_C}
  C =\frac{\beta}{\pi}\, \int_{0}^{2\pi/\omega_\text{B}} {\cal P}^{1}(t)\, \cos(\omega_\text{B} t)\, \id t \qquad \text{ as  }\; \omega_\text{B}\rightarrow 0 \,\ve\rightarrow 0
\end{equation}
From the combination of \eqref{eq_C} with \eqref{eq_P1} for the excess power ${\cal P}^1(t)$, we obtain the heat capacity. {\it A priori }we know that the dissipated power is measured by the area of the hysteresis loop.  Therefore, its response to a change in environment temperature is more problematic below the dynamical critical temperature that was introduced in Section \ref{hyst} --- see also Fig.~\ref{fig_beta_star}.  Nevertheless, the mathematical definition \eqref{eq_P1} makes sense and we have proceeded with the calculation for all temperatures.


\subsection{From the macroscopic evolution}\label{num}

\begin{figure}[hbt]
    \centering
    \includegraphics[width=0.6\textwidth]{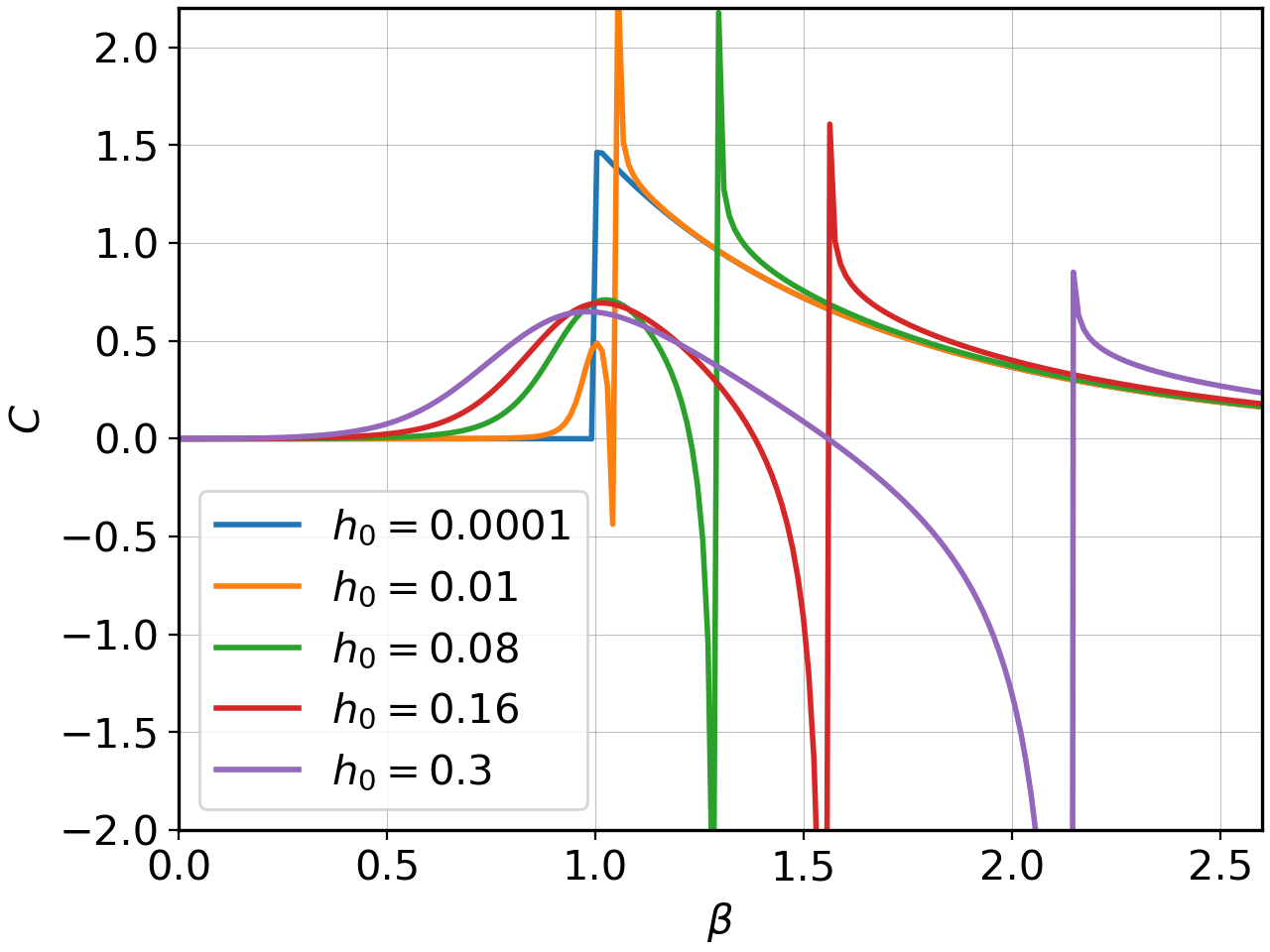}
    \caption{Specific heat as  function of $\beta$ obtained from the macroscopic evolution (unbounded rates in \eqref{bound}), for field amplitudes $h_0=0.0001,0.01,0.8,0.16, 0.3$ and with $\omega_0=0.02$ and $\omega_B=0.001$. These plots can be compared with the experimental result Fig.~3 in \cite{cera} for a rough resemblance.}
\label{fig_Cvsh0}
\end{figure}

We numerically solve \eqref{eq_P1} to obtain the excess power ${\cal P}^1(t)$.
Fig.~\ref{fig_Cvsh0} shows the heat capacity of a system with $\omega_0=0.02$ and $\omega_B=0.001$ for various values of the magnetic amplitude $h_0$. When $h_0 \to 0$, the expected equilibrium heat capacity is obtained. 

For $h_0 \neq 0$ the heat capacity approaches zero in both high and low--temperature regimes. In general, there is a local maximum around $\beta=1$ corresponding to the equilibrium critical temperature. Additionally, a novel behavior is observed here for the nonequilibrium case, where a divergence occurs at the dynamical critical temperature.
As $h_0$ increases (with other parameters held constant), the divergence becomes more pronounced, with its position shifting to the right, following $\beta_c(h_0,\omega_0)$ of Section \ref{hyst}. We also observe negative values for the heat capacity just before the divergence.

\begin{figure}[hbt]
    \centering
    \includegraphics[width=0.9\textwidth]{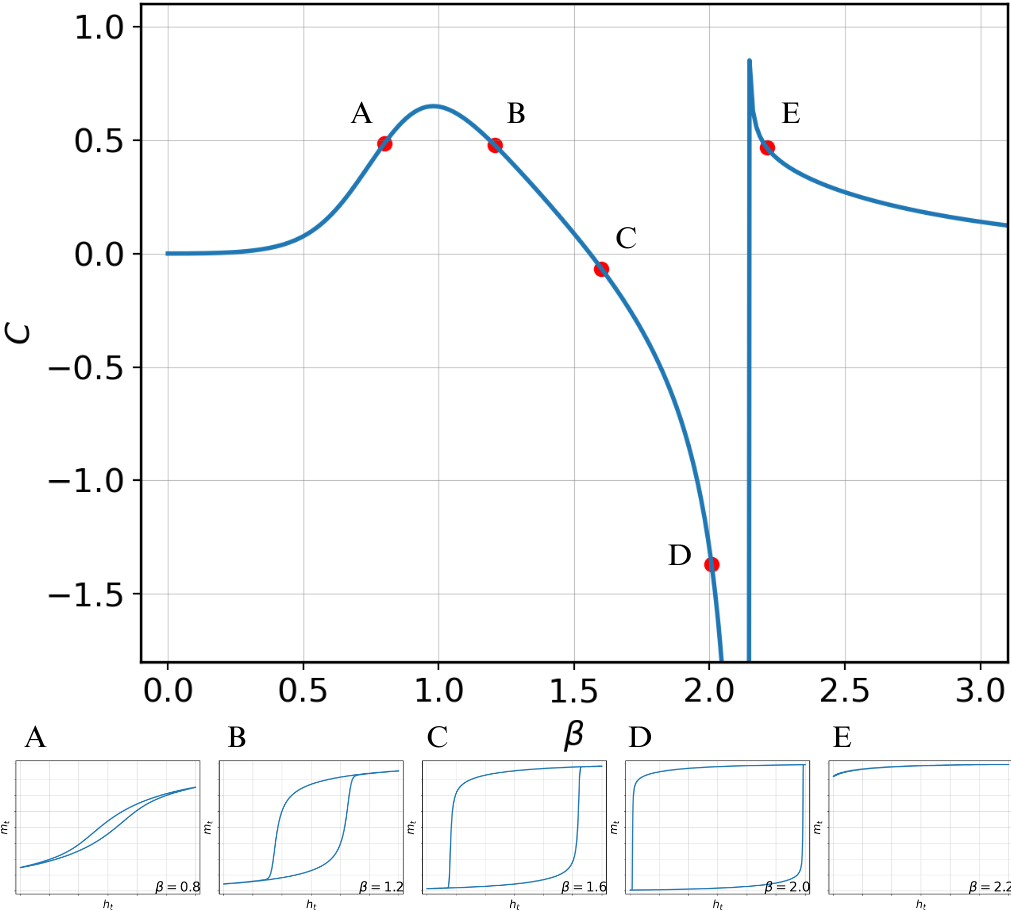}
    \caption{Specific heat capacity as function of $\beta$, obtained from the macroscopic evolution with unbounded rates \eqref{bound} and with $\omega_0=0.02$, $\omega_\text{B}=0.002$, $h_0=0.3$. }
\label{fig_Cloops}
\end{figure}

Fig.~\ref{fig_Cloops} shows the specific heat and magnetization at specific temperatures for $\omega_0=0.02$, $\omega_B=0.002$ and $h_0=0.3$. For these values, the divergence in the specific heat capacity occurs around $\beta_c\approx 2.15$ (dynamical critical temperature); see also Fig.~\ref{fig_fin_mag}. 
As the inverse temperature increases, the hysteresis loop widens, eventually exceeding $h_0$. Beyond this point, the magnetization becomes confined to either the positive or negative solution. This point defined the dynamical critical temperature in Section~\ref{hyst}, and is where the divergence in the (nonequilibrium) thermal response happens.\\

\begin{figure}[hbt]
    \centering
    \includegraphics[width=0.6\textwidth]{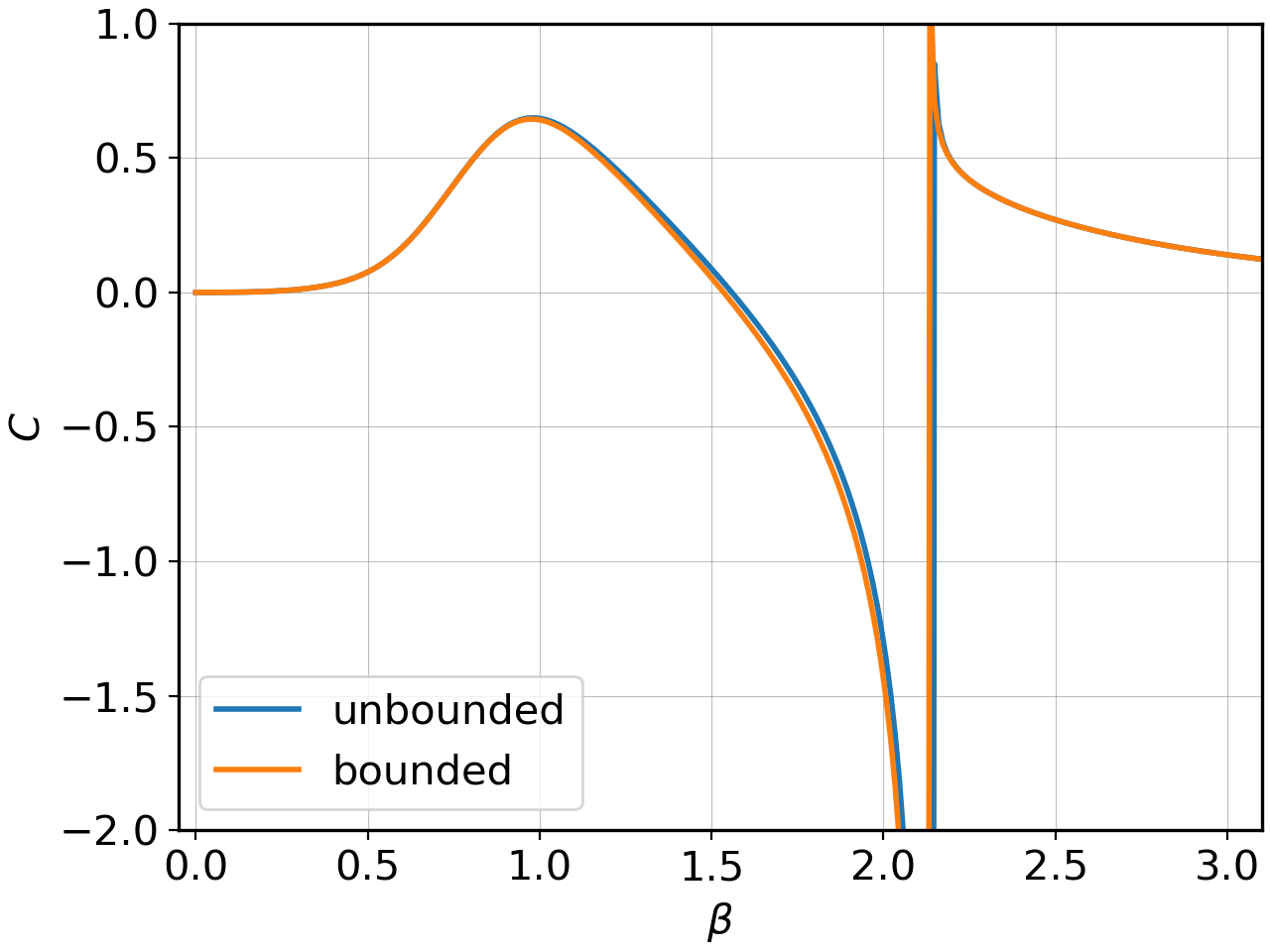}
    \caption{Specific heat capacity as function of $\beta$, obtained from the macroscopic evolution following \eqref{bound}, at $h_0=0.3$, $\omega_0=0.02$ and $\omega_\text{B} = \omega_0/20$.}
\label{bbuun}
\end{figure}

The case of bounded {\it versus} unbounded rates (see \eqref{bound}) is explored in Fig.~\ref{bbuun}, where little difference is noted.

\subsection{Finite--system simulation}\label{fss}

We use the Metropolis--Hastings algorithm for the simulation of the stochastic dynamics \eqref{eq:fin_trans_rate}; \cite{ROSS2023279}. 
To determine the excess power, ${\cal P}^1(t)$, we conduct two simulations: one with a time--dependent temperature and another with a constant temperature. The excess power is then obtained by calculating the difference $({\cal P}(t)-{\cal P}^0(t))/\ve$. The specific heat is calculated using \eqref{eq_C}.\\

\begin{figure}[hbt]
    \centering
    \includegraphics[width=0.7\textwidth]{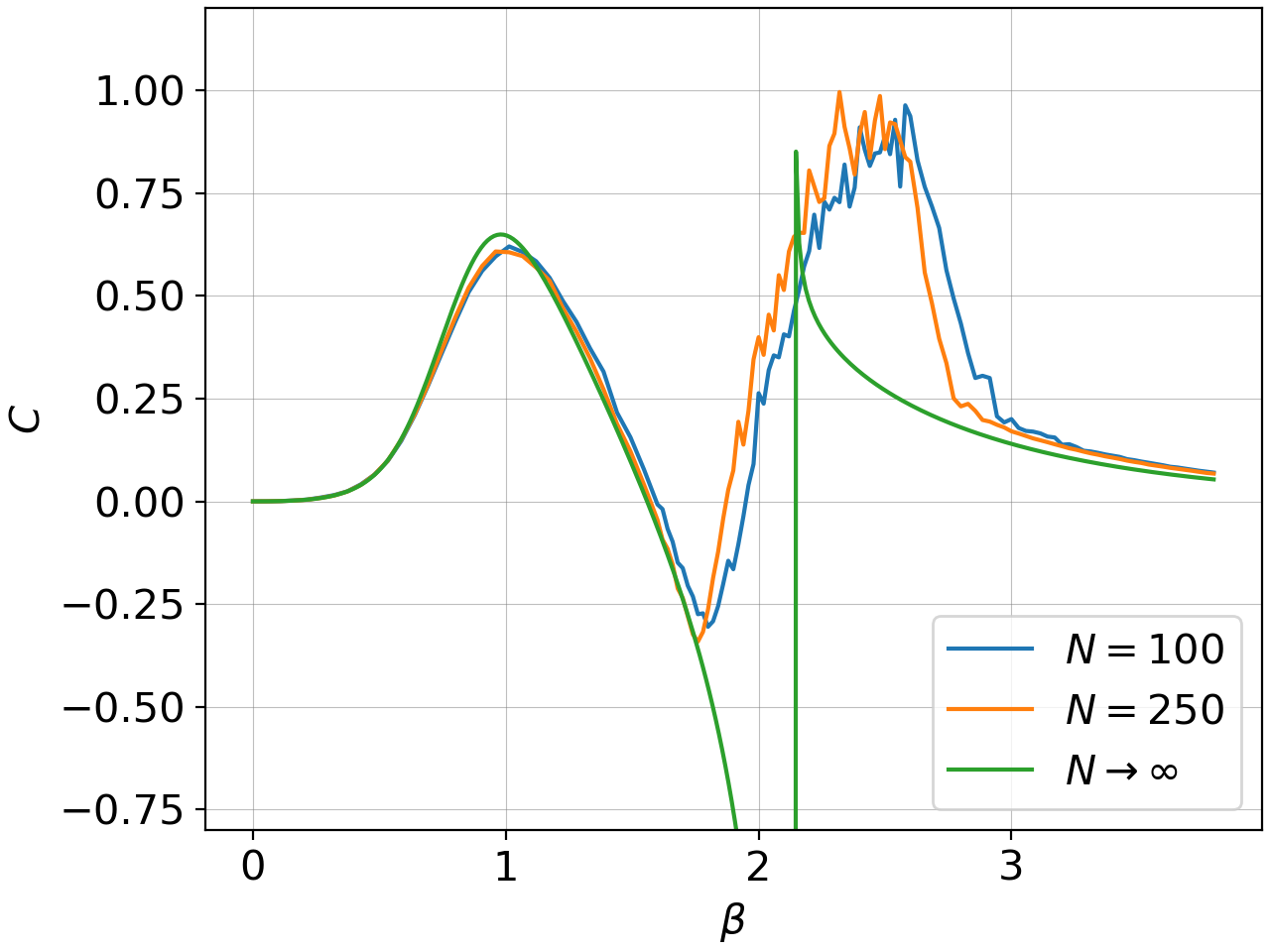}
    \caption{Specific heat as function of $\beta$ for a system simulation with $100$ and $250$ spins and for the macroscopic evolution (using bounded rates \eqref{bound} in all cases). The system has $h_0=0.3$ and $\omega_0 = 0.02 = 15\omega_B$. For the finite systems, $\epsilon = 0.2$ and the heat capacity was averaged over 2500 runs in the $N=100$ case, for 2000 runs in the $N=250$ case.}  
    \label{fig_finspin_allC}
\end{figure}

Fig.~\ref{fig_finspin_allC} shows the specific heat for systems with $N=100$ and $N=250$ spins, and compares it with the previous calculation from the macroscopic evolution.
Above the dynamical critical temperature, the results follow the macroscopic results very well. For lower temperatures, where the hysteresis loop disappears, the signal remains noisy.

\section{Conclusions}

Mean--field systems are essential in exploring critical behavior in spatially extended nonequilibrium systems. In this work, we study the thermal behavior of the Curie--Weiss spin system under the influence of a time--dependent magnetic field. We use the nonequilibrium heat capacity as the primary observable to characterize the system's response.

We find that the behavior of the heat capacity significantly differs from its equilibrium counterpart. A local maximum appears around the equilibrium transition temperature. More surprisingly, there is a region of negative heat capacity, followed by a sudden change in its sign and a subsequent monotonic decrease to zero for high inverse temperatures. The divergence of the specific heat occurs at the dynamical critical temperature as defined from the behavior of the time--dependent magnetization and its ability to form a hysteresis loop.

\vspace{1cm}
\noindent {\bf Acknowledgment:}  We are grateful to Christ Glorieux, Karel Netočný and Roi Holtzman for very useful discussions and suggestions.

\bibliographystyle{unsrt}  
\bibliography{chr}
\onecolumngrid

\end{document}